\documentclass[12pt]{article}
\usepackage{epsfig}
\newcommand{\be}{\begin{eqnarray}}
\newcommand{\ee}{\end{eqnarray}}
\renewcommand{\d}{\mathrm{d}}
\newcommand{\feynmandag}{\slash\hspace{-0.55em}}
\newcommand{\alphas}{\alpha_{\mathrm{s}}}
\newcommand{\xBj}{x_{\mathrm{Bj}}}
\newcommand{\gD}{g_{_D}}
\newcommand{\fDs}{f_{D_s}}
\newcommand{\Ds}{D_s^+}
\newcommand{\MDs}{M_{D_s}}
\newcommand{\mc}{m_c}
\newcommand{\ms}{m_s}
\begin{document}

\begin{center}
{\LARGE Diffractive $D_s$ production in charged current DIS}\\[0.8cm]
{\large Zhongzhi Song$~^{(a)}$ and~Kuang-Ta Chao$~^{(b,a)}$}\\[0.5cm]
{\footnotesize (a)~Department of Physics, Peking University,
 Beijing 100871, People's Republic of China}

{\footnotesize (b)~China Center of Advanced Science and Technology
(World Laboratory), Beijing 100080, People's Republic of China}
\end{center}

\vspace{0.5cm}
\begin{abstract}
We present a perturbative QCD calculation of diffractive $D_s$
production in charged current deep inelastic scattering. In the
two-gluon exchange model, we analyze the diffractive process $\nu
N \rightarrow \mu^- N \Ds$, which may provide useful information
for the gluon structure of nucleons and the diffraction mechanism
in QCD. The cross section of diffractive $D_s$ production with
$\xBj=0.005$-0.05 and $E_\nu=50$ Gev is found to be
$2.7\times10^{-5}$ pb. In spite of this small cross section, the
high luminosity available at the $\nu$-Factory in the future would
lead to a sizable number of diffraction events.


\end{abstract}
Diffractive leptoproduction of mesons has received much
attention\cite{e1}-\cite{e3}due to two reasons. First it is of
interest for the study of diffractive production mechanism within
QCD and second, its cross section is dominantly proportional to
the square of the gluon density in the nucleon, e.g., in the case
of diffractive $J/\psi$ electroproduction.

Aside from the diffractive electroproduction processes, the
charged current(CC) induced diffraction may also be interesting.
To the lowest order in perturbative QCD, CC diffractive deep
inelastic scattering (DIS) \cite{c1} proceeds by the
Cabibbo-favored production of the $(u\bar{d})\,$and\,$(c\bar{s})$
states, and the two-gluon exchange between the $(c\bar{s})$ and
the nucleon may be the dominant mechanism for the diffractive
production of charmed strange mesons. With the help of high
luminosity available at the $\nu$-Factory, neutrino induced
diffraction in CC DIS can shed more light on the QCD mechanism of
diffractive meson production. At the same time, it is a new way to
study the gluon structure of nucleons.

\begin{figure}
\centering
\includegraphics[width=6cm]{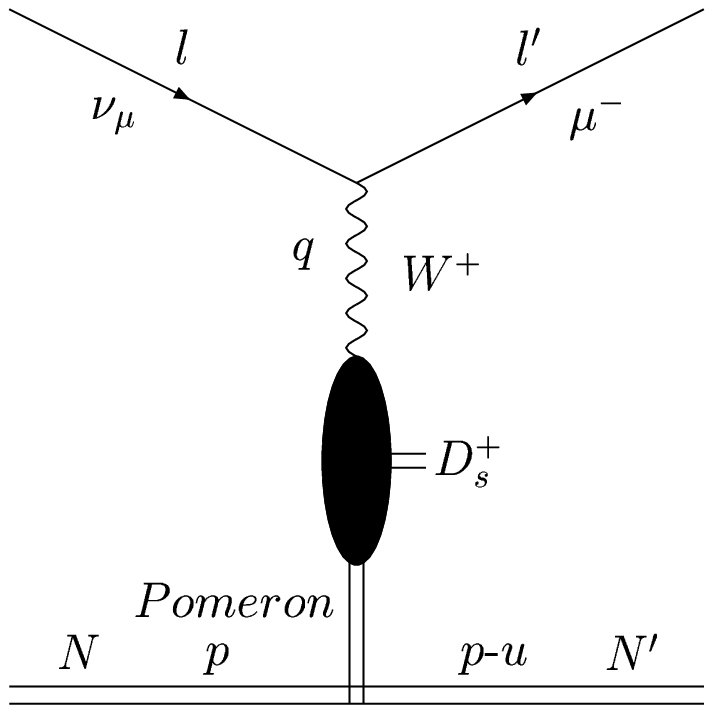}
 \caption{Diagram for the neutrino induced diffractive $\Ds$ production.}
 \label{gra1}
\end{figure}

We now consider the diffractive process(Fig.~\ref{gra1}\,),
 \be
 \nu_{\mu} + N \to \mu^- + N' + \Ds .
 \label{pro}
 \ee
We shall be concerned with the kinematic region where Bjorken
variable $\xBj=Q^2/2 p\cdot q$ is small. The three-fold
differential cross section is
 \be
 \frac{\d\sigma}{\d\xBj\d Q^2\d t} =
 \frac{e^2}{32(4\pi)^3\sin^2\theta_W}
 \frac{L_{\mu\nu}A^\nu A^{\mu*}}{\xBj s^2(Q^2+M_W^2)^2}
  \label{dif}
 \ee
where $s=(p+l)^2\approx 2 p\cdot l$, $t=u^2$ and $q^2=-Q^2$, $p$
and $l$ are the 4-momentum of the nucleon and the lepton
respectively. The leptonic tensor is
 \be
  L_{\mu\nu}=Tr[{\feynmandag l}'\gamma_\mu(1-\gamma_5)\feynmandag
  l\gamma_\nu].
 \label {lep}
 \ee

\begin{figure}
\centering \includegraphics[width=6cm]{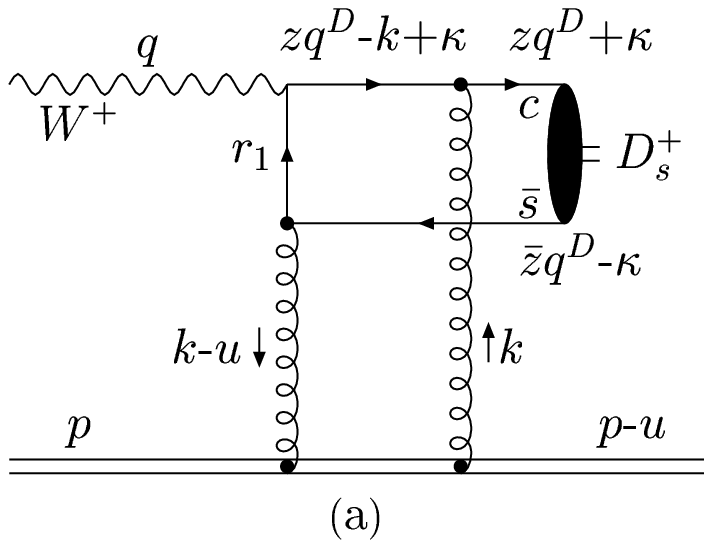}%
 \hspace {1cm}%
 \includegraphics[width=6cm]{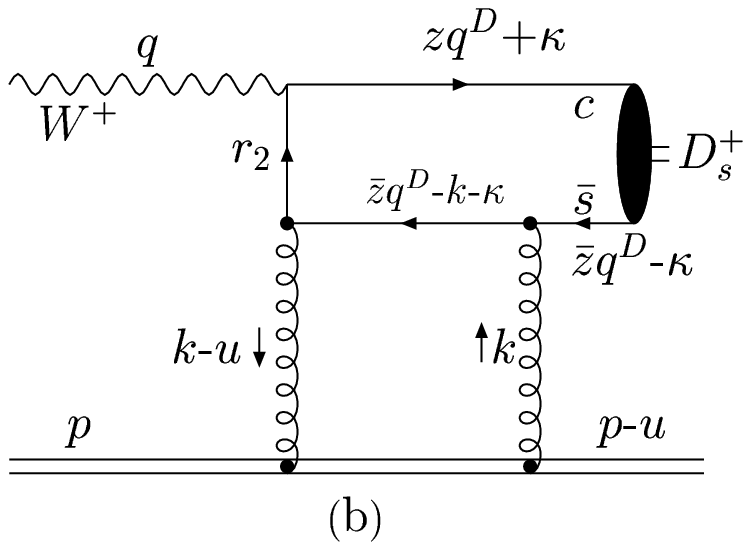}
 \caption{Two of the four sub-diagrams. The other two diagrams are
obtained by interchanging $c$ and $\bar s$ quark lines.}
 \label{gra2}
\end{figure}

To the lowest order in perturbative QCD, the hadronic current
$A_\mu$ can be calculated from the colorless two-gluon exchange
subprocesses shown in Fig.~\ref{gra2}. We will use the
nonrelativistic approximation writing the $\Ds$ vertex in the form
$\gD (\feynmandag q^D+\MDs)\gamma_5$. The constant $\gD$ specifies
the $c\bar s$ coupling to the $\Ds$. We choose the $\Ds$ wave
function as
 \be
  \Psi_{\Ds}(z,\kappa_T)=\delta^{(2)}\,(\kappa_T)\delta(z-\mc/\MDs)
  ,
 \label{wav}
 \ee
where $z$ and $\bar z =(1-z)$ denote the fractions of $\Ds$
momentum carried by the $c$ and $\bar s$ quarks respectively,
$\kappa$ is their relative momentum. Here we take $\MDs\approx\mc
+ \ms$.

We first evaluate the gluon loops in the Feynman diagrams shown in
Fig.~\ref{gra2}. It is convenient to perform the loop integration
in terms of Sudakov variables\cite{e3}. That is, for all particles
the 4-momentum are decomposed in the form \be
 k_i=\alpha_i q' + \beta_i p' +k_i{_T} ,
 \label {sudakov}
 \ee
where $p'$ and $q'$ are respectively the light-like momenta of the
nucleon and $W^+$ boson, that is, $p'^2=q'^2=0$. In particular
 \be
 p=p'+\alpha_{_N} q', \hspace {2cm}  q=q'+\beta_{_W} p' ,
 \label{decom}
 \ee
with $ \alpha_{_N}=m^2_{_N}/2 p'\cdot q'$ and $\beta _{_W} =-Q^2/2
p'\cdot q'$. We consider the limit $2p'\cdot q' \gg m^2_{_N},
Q^2$, then we have $2p'\cdot q'\approx 2p\cdot q$.

Within the nonrelativistic approximation the quarks with
momenta(see Fig.~\ref{gra2}\,) $ z q^D+\kappa$ and $\bar z q^D
-\kappa$ are almost on mass shell. The integration over the gluon
longitudinal momentum leads, in the first diagram, the upper quark
with momentum $z q^D-k+\kappa$ to be on shell, leaving only the
quark propagator $(r^2_1 -\ms^2)^{-1}$ to be integrated over in
the gluon $k_{_T}$ integration.

Using the Sudakov decomposition, we find
 \be
 r^2_1-\ms^2=-\frac {1}{z}\,({\overline Q}^2+k^2_T)
 \label {propa}
 \ee
where $ {\overline Q}^2=z(1-z)(Q^2+\MDs^2)$, which is the relevant
effective perturbative QCD factorization scale\cite{scale}.

Taking the CKM matrix element $V_{cs}=1$, we write the
contribution given by the Feynman graph in Fig.~\ref{gra2}a as
 \be
  A^\mu_1 =
  \frac{i e \gD g^4_s F_c }{8\sqrt{2}z \pi \sin\theta _W }\int
  \frac {d k^2_T Tr{[\gamma_5({\feynmandag q^D}+\MDs){\feynmandag p}
(z{\feynmandag q^D}-{\feynmandag k}+\mc)\gamma ^\mu (1-\gamma_5)
({\feynmandag r_1}+\ms){\feynmandag p}}] \phi (k)}
  {(2p\cdot q)k^2 (k-u)^2 (r^2_1-\ms^2)}
 \label{ampli}
 \ee
where $F_c=2/3$ is the color factor, and $\phi (k)$ describes the
emission of the gluon pair by the proton\cite{e3},
 \be
 \phi (k)=\frac {3\pi}{4\alpha_s} f_{BFKL} (x, k^2_T).
  \label{emiss}
  \ee
where $f_{_{BFKL}}$ is the gluon density unintegrated over
$k_{_T}$ that satisfies the BFKL equation which effectively resums
the leading $\alpha _s {\ln [1/x]}$ contributions, with
\be
x\approx\beta_{D_s}-\beta_{_W}=\frac{Q^2+\MDs^2}{2p\cdot q}.
 \ee

To relate $f_{_{BFKL}}$ to the conventional gluon density, which
satisfies GLAP evolution, we must integrate over $k^2_{_T}$
\be
 xg(x,{\overline Q}^2)=
 \int ^{{\overline Q}^2} \frac {\d k^2_{_T}}{k^2_{_T}} f_{_{BFKL}}(x,
 k^2_{_T}).
  \ee

In analogy to the derivation of (\ref{ampli}), we find the sum of
the four diagrams in Fig.~\ref{gra2} is
 \be
 A_\mu=\sum^4_{i=1} A^i_\mu=
 \frac {3 \pi i z (1-z) e \gD g^2_s F_c (2 p\cdot q)}{\sqrt{2} \sin {\theta_W}}
\int \frac { {\d k^2_T} \hspace {5mm} q^D_\mu}
 {{\overline Q}^2{({\overline Q}^2+k^2_T)}}
 \frac {\partial (xg(x,k^2_T))} {\partial k^2_T}.
\ee

To the lowest order in $k^2_T$, we have \be
 A_\mu=\frac {3 \pi i z (1-z) e \gD g^2_s F_c (2 p\cdot q) q^D_\mu}
 {\sqrt{2} \sin {\theta_W}{\overline Q}^4}( xg(x,{\overline Q
 }^2)).
 \ee

So far we have calculated only the imaginary part of the
amplitude. We can use dispersion relations \cite{e2,collions} to
determine the real part, and numerically we find it to be not
negligible. Including the real part contribution as a perturbation
we now rewrite  the differential cross section (\ref{dif}) as
 \be
\frac{\d\sigma}{\d\xBj\d Q^2\d t} =
 \frac{e^2(s-2p\cdot q)|A|^2}{16(4\pi)^3\sin^2\theta_W\xBj
 s(Q^2+M_W^2)^2},
\ee
where
 \be
 A=\frac {12 \pi^2 i e \gD \alphas({\overline Q}^2) F_c }{\sqrt{2} \sin{\theta_W}{\overline Q}^2}
 [ xg(x,{\overline Q }^2)+\frac{i\pi}{2}\frac{\partial( xg(x,{\overline Q }
^2))}{\partial \ln x}]. \ee

Since we are concerned with small $x$, the effect of the nonzero
value of $|t|$, of which the minimum is $x^2 m^2_{_N}$, is
expected to be small. Then we can integrate out $t$ by \be
 \int \d t\,e^{-bt}=\frac{1}{b}
 \ee
where we will use the experimental slope value $b=3.3\,\mathrm
{Gev}^{-2}$ as in similar processes\cite{exp1}.
\begin{figure}
\centering \includegraphics[width=6cm]{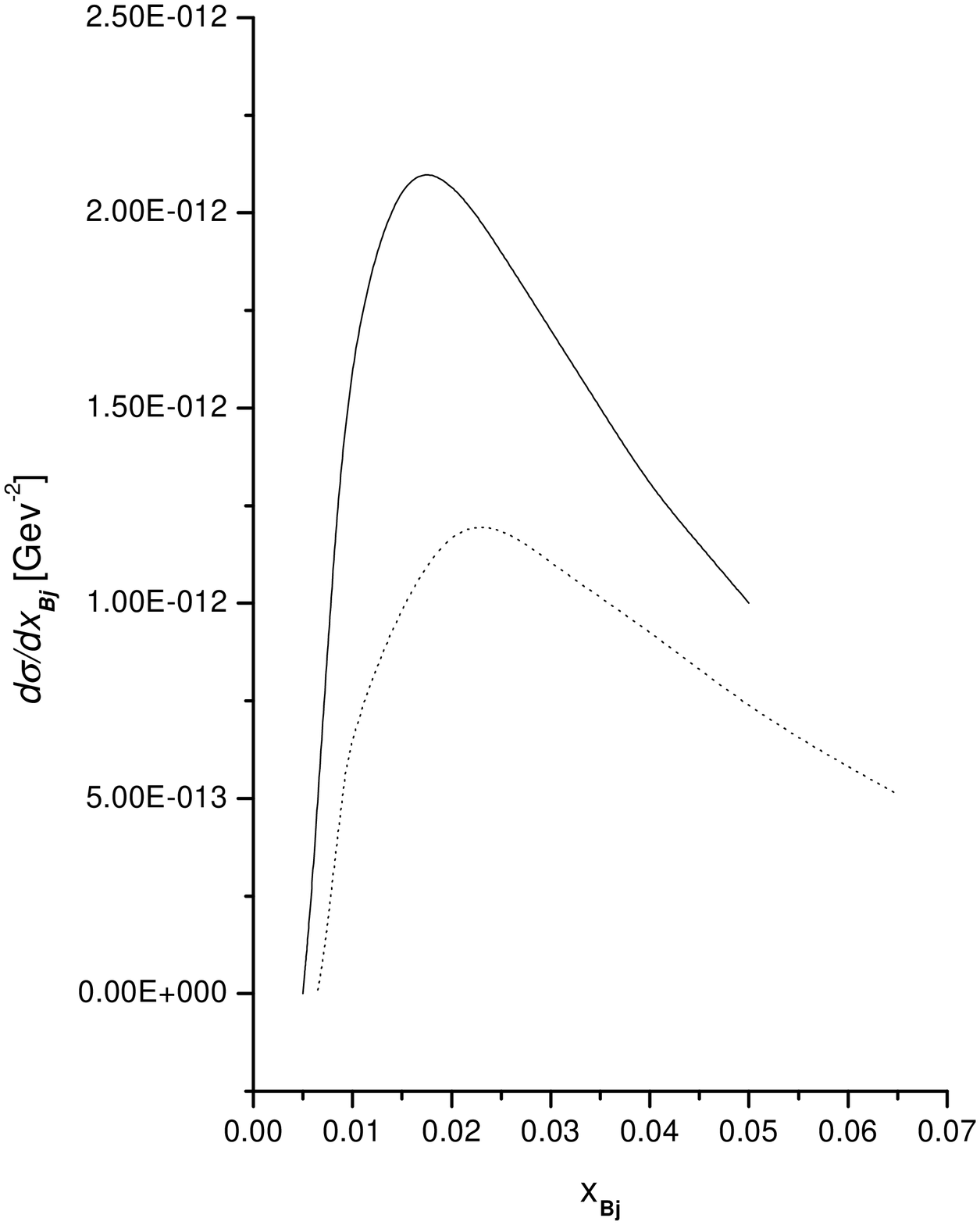}%
 \hspace {0.5cm}%
 \includegraphics[width=6cm]{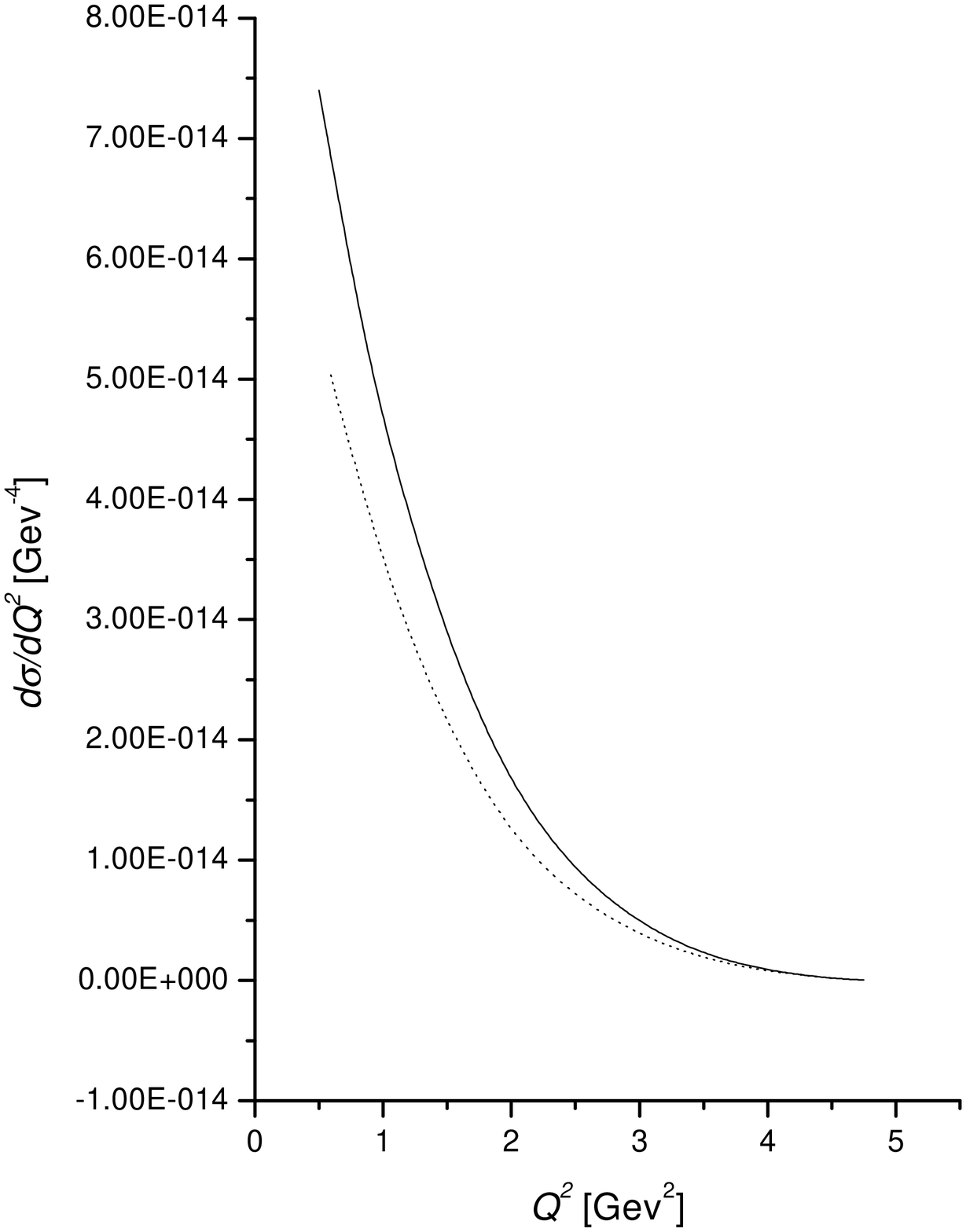}
 \caption{Differential cross sections as a function of $\xBj$ and $Q^2$.
 The neutrino energy has been chosen as $E_\nu=50$ Gev (solid lines)
  and $E_\nu=40$ Gev (dotted lines).}
 \label{gra3}
\end{figure}

To give numerical results, we take the input parameters as
follows: $M_W=80.4$ Gev, $\mc=1.5$ Gev, $\ms=0.5$ Gev. The running
strong coupling constant is chosen with $\alphas(\mc^2)=0.27$. For
the gluon distribution function, we select the
Gl\"{u}ck-Reya-Vogt(GRV) next-to-leading order(NLO) set\cite{grv}.
The constant $\gD$ can be expressed in terms of the decay constant
by
 \be
   \langle 0|\bar c\gamma_\mu (1-\gamma_5) s|\Ds\rangle=\fDs
   q^D_\mu ,
    \ee
which gives $\gD=\fDs/4$. Here we choose $\fDs=280$ Mev\cite{fds}.

In Fig.~\ref{gra3}(solid lines) we show the results obtained for
the differential cross sections $\d\sigma/\d\xBj$ and $\d\sigma/\d
Q^2$. The neutrino energy has been chosen as $E_\nu=50$ Gev. For
the plot of $\xBj$-dependence, $Q^2$ has been integrated from 0.5
Gev$^2$ to the upper bound given by the constraint on inelasticity
$y=Q^2/2\xBj p\cdot l < 1$. In the plot of $Q^2$-dependence,
$\xBj$ has been integrated from lower bound to 0.05 and taking the
same kinematic constraint mentioned above. Integrating over $Q^2$
and $\xBj$ in the kinematical region specified above gives a value
for the total cross section of $ \sigma=2.7\times10^{-5}$ pb.

To see the sensitivity of the differential cross sections to the
neutrino energy, we also present the results for $E_\nu=40$ Gev in
Fig.~\ref{gra3}(dotted lines). The kinematic regions of $Q^2$ and
$\xBj$ are the same as in the $E_\nu=50$ Gev case except that the
upper bound of $\xBj$ is chosen as 0.065. Integrating out all
variables gives the total cross section of
$\sigma=2.0\times10^{-5}$ pb.  In spite of the small cross
section, the high luminosity available at the $\nu$-factory in the
future \cite {nufact} would lead to a sizable number of events of
the order of magnitude $10^4$.

Some discussions are in order. First, in the two gluons exchange
processes in general we should encounter the  so-called
off-diagonal gluon distribution function \cite {ji}. But it is
expected that for small x there is no big difference between the
off-diagonal and the usual diagonal gluon densities \cite {hood}.
So in the above calculations we have estimated the small x
production rate by approximating the off-diagonal gluon density by
the usual gluon density. This situation is similar to many
diffractive production processes at hadron colliders {\cite
{yuan1,yuan2}}(for detailed discussions, see \cite {yuan1}).

Second, we have used $ {\overline Q}^2=z(1-z)(Q^2+\MDs^2)$ as the
energy scale for the application of the perturbative QCD. The
applicability of pQCD is guaranteed by the large value of
${\MDs}^2\approx(\mc + \ms)^2$. So $Q^2$ can be chosen to be
rather small, say, 0.5 $\sim$ 1.0 Gev$^2$.

Third, we have used nonrelativistic approximation to describe the
$D_s$ wavefunction, and this will cause some uncertainties in our
calculation. Relativistic effects can be quite important and
should be further considered in a similar way as in {\cite
{e3,tang}}.

In conclusion, we have calculated the diffractive $D_s$ production
rate in the neutrino induced charge current DIS process in the
two-gluon exchange model in QCD, and found the diffractive
production of $D_s$ to be observable with the high luminosity
available at the $\nu$-Factory in the future.

After the calculation in this work was completed, B.
Lehmann-Dronke and A. Sch\"{a}fer \cite {schafer} published a
preprint treating a similar process to that we considered. But
they analyzed exclusive $D_s$ production in the large $\xBj$
region, whereas we studied the diffractive production of the $D_s$
meson in the two-gluon exchange model with small $\xBj$.  Although
calculated in different methods and in different kinematic
regions, our total cross section has the same magnitude as theirs.

We would like to thank F. Yuan for his valuable discussions and
K.Y. Liu for some numerical calculations. This work was supported
in part by the National Natural Science Foundation of P.R. China,
and the Education Ministry of P.R. China.

\end{document}